\documentclass[utf8]{FrontiersinVancouver} 

\usepackage{url,hyperref,lineno,microtype,subcaption}
\usepackage[onehalfspacing]{setspace}
\usepackage{makecell}
\usepackage{booktabs}
\usepackage{multirow}
\usepackage{xcolor}
\usepackage{url}

\def\keyFont{\fontsize{8}{11}\helveticabold }
\def\firstAuthorLast{Osaba {et~al.}}
\def\Authors{Eneko Osaba\,$^{1*}$ and Esther Villar-Rodriguez\,$^{1}$}
\def\Address{$^{1}$TECNALIA, Basque Research and Technology Alliance (BRTA), 48160 Derio, Spain}
\def\corrAuthor{eneko.osaba@tecnalia.com}
\def\corrEmail{}

\begin{document}
\onecolumn
\firstpage{1}

\title[Transfer of Knowledge through Reverse Annealing]{Transfer of Knowledge through Reverse Annealing: A Preliminary Analysis of the Benefits and What to Share} 

\author[\firstAuthorLast ]{\Authors}
\address{}
\correspondance{}

\extraAuth{}

\maketitle

\begin{abstract}
Being immersed in the NISQ-era, current quantum annealers present limitations for solving optimization problems efficiently. To mitigate these limitations, D-Wave Systems developed a mechanism called Reverse Annealing, a specific type of quantum annealing designed to perform local refinement of good states found elsewhere. Despite the research activity around Reverse Annealing, none has theorized about the possible benefits related to the transfer of knowledge under this paradigm. This work moves in that direction and is driven by experimentation focused on answering two key research questions: \textit{i)} \textit{is reverse annealing a paradigm that can benefit from knowledge transfer between similar problems?} and \textit{ii)} \textit{can we infer the characteristics that an input solution should meet to help increase the probability of success?} To properly guide the tests in this paper, the well-known Knapsack Problem has been chosen for benchmarking purposes, using a total of 34 instances composed of 14 and 16 items.
\tiny
 \keyFont{ \section{Keywords:} Quantum Annealing, Reverse Annealing, D-Wave, Quantum Optimization, Transfer Optimization}
\end{abstract}

\section{Introduction}

A quantum annealer (QA, \cite{rajak2023quantum}) is a specific kind of quantum device designed to deal with optimization problems by means of a process inspired by classical Simulated Annealing \cite{bertsimas1993simulated}. These computers leverage quantum mechanics to efficiently explore solution spaces in an attempt to find the optimum value of a given objective function. Today, advances in quantum technologies have contributed to the building of intermediate-scale QAs that implement quantum annealing for programmable use. There are different platforms for building quantum annealers, such as optical tweezers \cite{ebadi2022quantum} or superconducting integrated circuits \cite{bunyk2014architectural}, with the latter being the most recognized to date. Additionally, several companies are working on this technology and building their own devices, such as NEC\footnote{https://parityqc.com/a-new-quantum-annealer-by-nec}, Qilimanjaro\footnote{https://qilimanjaro.tech/}, and D-Wave Systems\footnote{https://www.dwavesys.com/}. 

However, all the progress made in recent years was born in the NISQ-era \cite{preskill2018quantum}, when quantum devices present great limitations in solving optimization problems efficiently, even when these problems are small- or medium-sized. As a consequence, the community as a whole is striving to devise schemes and mechanisms to address the current limitations and take advantage of the potential that quantum computing currently offers. Among the most common strategies are the design and implementation of advanced hybrid resolution schemes, which leverage the best of both computing paradigms \cite{osaba2024hybrid}. Other alternatives include the implementation of error mitigation strategies \cite{bhalgamiya2022quantum} or making use of reliable quantum simulators \cite{celeri2023digital}.

As part of these activities, an interesting strategy studied in works such as \cite{perdomo2011study,grass2019quantum,ohkuwa2018reverse} involves initiating the quantum evolution with a state that is an educated guess of the problem’s solution. It is interesting to note that in gate-based quantum computing, similar strategies have also been explored, with the warm-start initialization of QAOA algorithms being particularly well-known \cite{tate2023bridging,beaulieu2021max,tate2023warm}.

In this paper, we focus on a specific strategy called Reverse Annealing (RA, \cite{DWaveRA}), which was introduced by D-Wave Systems in 2017. The motivation behind the implementation of RA lies in the limitations of D-Wave’s annealers to conduct local refinement of states previously found either through a forward anneal or by means of a classical technique. Thus, RA allows quantum local search by annealing backward from a previously specified state, and then forward to a new one.

Today, there are many practical studies that employ RA for improving a solution previously found by other solving schemes \cite{haba2022travel,carugno2022evaluating}. A significant amount of research has also been published studying theoretical aspects of RA, such as its sensitivity to proper parameterization, mainly regarding the \texttt{RA Schedule} \cite{grant2021benchmarking,golden2021reverse}. Despite the existence of this buoyant research activity, the full potential of RA is yet to be revealed, as some crucial aspects of its performance still require a deeper analysis.

On this basis, and bringing to the fore the proven potential of RA, this paper is committed to extending the frontiers of this promising mechanism, exploring its application in situations not yet studied by the community. To achieve this, the analysis provided in this work is driven by experimentation focused on answering two key research questions (RQ).

As mentioned before, RA was conceived as a local refinement mechanism. For this reason, in order to solve a problem using this paradigm, the QA must be initialized with a known (classical) solution. RA explores the local space around that solution for bitstrings with even lower energy. Despite the existence of intense research activity around RA, none has theorized about the possible benefits related to the transfer of knowledge (ToK, \cite{gupta2017insights}) under the umbrella of this paradigm. So far, all studies feed the annealer with an existing solution to the same problem, with this existing solution usually obtained either by a forward annealing process executed just before the RA \cite{pelofske2023initial} or by a classical optimization method \cite{venturelli2019reverse}. However, it would be reasonable to think that, in a real application context where the problems faced by the industry share analogies, a previous solution may be a good starting point for the resolution of a new (or perhaps not so new, as we argue) problem. At the time of writing, the use of solutions coming from different problems has remained unexplored. In other words, the influence of solving a specific problem through the RA paradigm by using as input a solution from another similar but not exactly the same problem has not been studied. In this context, the first of the research questions is:

\begin{quote}
	\textit{\textbf{RQ1}: Is RA a paradigm that can benefit from knowledge transfer between problem instances with similar characteristics?}
\end{quote}

In relation to the analysis of RA behavior, several studies have tried to provide insights about the requirements that input solutions have to meet, determining in many cases that RA is efficient when it is fed by a solution \textit{close} enough to the optimum of the problem to be solved. However, the term \textit{close} is usually ambiguously used \cite{ding2024effective}, with studies translating this closeness into energy \cite{passarelli2020reverse}, while others into the composition of the solution itself, i.e., the difference in terms of the Hamming distance between the input solution and the optimum one \cite{pelofske2023mapping}. Thus, the second RQ posed in this paper aims to shed light on the characteristics that an input solution should have to increase the probability of success:

\begin{quote}
	\item \textit{\textbf{RQ2}: Can we infer the characteristics that an input solution should meet to help increase the probability of succeeding in a RA process?}     
\end{quote}

Moving in this direction is especially interesting for ToK since it helps to identify which solution or solutions from the sampling pool are the most promising to feed RA and to set the analysis framework for future work in this area. For clarification purposes and following the nomenclature of ToK, the term \textit{source} refers to the problem or task whose knowledge will be leveraged to solve a \textit{target} task \cite{villar2025transfer}.

For properly guiding the experimentation of this paper, the well-known Knapsack Problem (KP, \cite{salkin1975knapsack}) has been used as a benchmarking problem. We have chosen KP because:

\begin{itemize}
	\item It has been extensively used for benchmarking purposes in QC-oriented studies \cite{bozejko2024optimal,wilkening2023quantum,osaba2025d}.
	\item It is an appropriate problem to be formulated as a QUBO \cite{glover2018tutorial}.
	\item It is a complex problem to be solved by quantum algorithms, as has been previously demonstrated in the literature \cite{pusey2020adiabatic,van2021quantum}.
\end{itemize}

Two main instances have been used in the planned experimentation, composed of 14 and 16 items, respectively. We call these seed instances \textit{parent}-instances. From each of them, 16 new cases with similar characteristics have been created, hereinafter named \textit{descendant}-instances. With all this, the main objective of the experimentation is to analyze, given a target-instance, the impact of feeding RA with solutions obtained from similar yet different tasks.

It is worth noting that the size of the instances has not been chosen randomly. This selection has been made after thoroughly analyzing the current state of the art in the conjunction between QA and KP. Additionally, our motivation has been to construct instances that are, on one hand, large enough so that D-Wave’s QA would not always solve them optimally through the forward annealing process, and on the other hand, small enough to be efficiently embedded in the quantum computer.

Lastly, we have used the Hamming distance as a reference measure to evaluate the difference between two solutions. This measure stands out particularly for its simplicity and efficiency in calculation. Additionally, the Hamming distance is especially suited for measuring the similarity among binary strings, which are often used in quantum computing. It has also been frequently used for comparing DNA sequences and error-correction codes.

\section{Method} \label{sec:RA}

As mentioned in the introduction, a QA is a specialized quantum device engineered to tackle optimization problems through a process inspired by classical Simulated Annealing. More specifically, a QA operates on the principle of adiabatic computation, where an initial, easily prepared Hamiltonian is gradually evolved from its ground state to the ground state of a final, problem-specific Hamiltonian. If this evolution is sufficiently slow, the adiabatic theorem ensures that the system stays in the ground state throughout the computation. In QA, the adiabatic theorem is deliberately relaxed, permiting the process to evolve more quickly than the adiabatic limit would normally permit. Consequently, transitions to higher energy states often occur during the evolution. Despite this computational model can also be considered also universal \cite{mizel2007simple}, the D-Wave annealer, which is the one used in this study, is based on an Ising Hamiltonian, which limits the kind of problems that can be executed on the computer. Nevertheless, this type of QA is well-suited for solving combinatorial optimization problems as the one addressed in this paper \cite{yang2023survey}.

Regarding the RA method, it is a specific type of quantum annealing process designed to perform local refinement of good states found elsewhere \cite{DWaveRA}. To achieve this, RA resorts to a time-dependent Hamiltonian:

\begin{equation}\label{ref:H}
	H(t)=A(s(t))H_0 + B(s(t))H_1
\end{equation}
in which $s(t)\in[0,1]$, and $A(s)$ and $B(s)$ are time-dependent amplitudes that must satisfy $A(0) \gg B(0)$ and $A(1) \ll B(1)$. Additionally, $H_0$ defines the initial Hamiltonian, while the final Hamiltonian $H_1$ corresponds to the unconstrained optimization problem, with a ground state that represents the computational solution. In contrast to forward annealing, RA reverses the time-evolution procedure by starting in an eigenstate of $H_1$ and evolving under Equation (\ref{ref:H}) in the opposite direction. Thus, $H(t) = H(1)$ at $t$=0, and the process is divided into three different steps in which the Hamiltonian:

\begin{enumerate}
	\item First evolves backward to a previously specified point $s_p$ in the control schedule in a time $t_r=t_1$ coined \textit{ramp-time}.
	\item Then is paused for a time $t_p=t_2-t_1$ (this step is optional).
	\item It finally evolves forward from $s_p$ at $t_2$ to the final Hamiltonian at time $T$ for a $t_q = T-t_2$ \textit{quench time}.
\end{enumerate}

Therefore, the \texttt{RA} \texttt{Schedule} can be defined as shown in the following Equation \ref{ref:RA} \cite{grant2021benchmarking}:

\begin{equation}\label{ref:RA}
	s(t) = \begin{cases}
		1+\frac{s_p-1}{t_1}t, & 0\le t \le t_1, \\
		s_p, & t_1\le t \le t_2, \\
		s_p+\frac{1-s_p}{T-t_2}(t-t_2), & t_2\le t \le T.
	\end{cases}
\end{equation}

\section{Experimentation \& Discussion} \label{sec:exp}

Before starting with the description of the experimentation, it is important to note that there is no metric that accurately indicates how similar two problems are to each other (in order to make them good candidates for ToK). The existence of such a metric would be very advantageous for this purpose. More specifically, similarity should be measured in terms of overlap in the energy landscape of a unified search space \cite{gupta2015multifactorial}. However, in the real world, it is not possible to determine this overlap beforehand without having executed and, therefore, solved the problem. Given this circumstance, similarity becomes a matter of intuition and subject to the domain knowledge acquired by the expert. This situation does not detract an iota of value because, in real industrial contexts, most problems have a repetitive nature: a pool of recurring customers in routing problems, production of similar materials with stable task typologies and machines, etc. In such situations, ToK is a promising strategy to resort to past solutions in order to improve the results or speed up the computation of a new, yet similar, task.

With this in mind, the experimentation carried out in this paper uses the well-known KP as a benchmarking problem. In a nutshell, KP consists of a set $P$ composed of $n$ items, describing each item $p_i$ by profit ($v_i$) and weight ($w_i$), which must be packed into a container with a maximum capacity $W$. The objective, therefore, is to select a subset of items to be stored that, without exceeding $W$, maximizes the profit obtained. It is worth noting that the metric used in this study for measuring the quality of a solution corresponds to the energy provided by the quantum annealer, which should be minimized. That is, the less energy, the better the solution.

Aiming to provide an answer to the above-presented \textbf{RQ}s, we have designed two separate experiments focused on two \textit{parent}-instances of the KP, which are composed of 14 (\texttt{s14}) and 16 (\texttt{s16}) items, respectively. Both instances have been generated ad hoc for this study with the values $v$ and $w$ randomly selected from \{1,2,3,4\}. Finally, $W=(\sum^n_{i=1} w_i)/2$.  

Taking these cases, each \textit{descendant}-instance has been generated by applying the following strategy: 
\begin{itemize}
	\item First, the newly generated \textit{descendant}-instance starts out as an exact copy of the corresponding \textit{parent}-instance.
	\item Then, a set $P'$ is created, consisting of all the unique items in $P$ along with their energy impact. Provided that each item $p_i$ in $P$ is described as a tuple ${v_i,w_i}$, a unique item in $P'$ means that it differs from the others in at least one element of the tuple.
	\item Next, a certain number of items of the \textit{descendant}-instance are modified following this criterion: when the purpose is to increase the energy impact of an item $p_i$, it is replaced by the next highest energetic item in $P'$. Analogously, $p_i$ is replaced by the next lowest energy item in $P'$ when the objective is to reduce the energy. It should be noted that the items with the highest and lowest energy are never modified in the creation of \textit{descendant}-instances. 
\end{itemize} 


Following this strategy, 16 \textit{descendant}-instances have been created for each \textit{parent}-instance, equally divided into four categories:

\begin{itemize}
	\item \texttt{X\_L2L} - \textit{Lowest-Energy-to-Lower-Energy}: X\% of the less energetic variables (i.e., items) of the \textit{parent}-instance are selected and modified so that their energetic impact is even lower.
	\item \texttt{X\_L2H} - \textit{Lowest-Energy-to-Higher-Energy}: X\% of the less energetic variables of the \textit{parent}-instance are selected and modified so that their energetic impact is greater.
	\item \texttt{X\_H2L} - \textit{Highest-Energy-to-Lower-Energy}: X\% of the highest energetic items of the \textit{parent}-instance are selected and modified so that their energetic impact is lower.
	\item \texttt{X\_H2H} - \textit{Highest-Energy-to-Higher-Energy}: X\% of the highest energetic variables are selected and modified so that their energetic impact is even greater.
\end{itemize}

It is noteworthy that solutions to these problems are represented by a $(n+s)$-long vector, where $n$ is the number of items that compose the instance 
and $s$ are slack terms introduced by the penalties. For both \texttt{s14} and \texttt{s16}, $s=5$, so that the decision variables for each problem are 19 and 21, respectively. In our study, the variables eventually modified correspond only to the first $n$ items. 


\begin{table}[!t]
	\centering
	\caption{Instances used in the experimentation. \textit{Energy wrt. \texttt{sXX}} represent the energy value of the best-found solution for the descendant-instances, using as a base the objective function of \texttt{sXX}. Hamming distances are calculated using as reference the best solution found for the \textit{parent}-instances and the \textit{descendant}-instances along 10 independent runs. Hamming distances are represented by the total value ($h$) and divided into variables that represent items ($n_h$) and slack variables ($s_h$).}
	\resizebox{0.85\columnwidth}{!}{
		\begin{tabular}{c|ccc|c|ccc}
			\toprule[1.4pt]                
			\textbf{Instance} & \makecell[c]{\textbf{Best}\\\textbf{energy}} & \makecell[c]{\textbf{Energy}\\\textbf{wrt. \texttt{s14}}} & \makecell[c]{\textbf{Hamming}\\\textbf{distance}} & \textbf{Instance} & \makecell[c]{\textbf{Best}\\\textbf{energy}} & \makecell[c]{\textbf{Energy}\\\textbf{wrt. \texttt{s16}}} & \makecell[c]{\textbf{Hamming}\\\textbf{distance}}\\
			\toprule[1.0pt]
			s14 & -12422 & -- & -- & s16 & -10716 & --  & -- \\ \midrule
			s14\_0.2\_L2L & -12019 & -12419 (3) & 7 (7, 0) & s16\_0.2\_L2L & -10133 & -10712 (4) & 9 (7, 2) \\
			s14\_0.2\_L2H & -13222 & -12421 (1) & 3 (3, 0) & s16\_0.2\_L2H & -11583 & -10713 (3) & 5 (4, 1)\\
			s14\_0.2\_H2L & -12021 & -12391 (31) & 5 (5, 0) & s16\_0.2\_H2L & -10137 & -10715 (1) & 4 (3, 1)\\
			s14\_0.2\_H2H & -12820 & -12388 (34) & 4 (4, 0) & s16\_0.2\_H2H & -10995 & -10668 (48) & 12 (11, 1)\\
			\midrule
			s14\_0.4\_L2L & -12820 & -12388 (34) & 8 (8, 0) & s16\_0.4\_L2L & -8974 & -10714 (2) & 5 (4, 1)\\
			s14\_0.4\_L2H & -13620 & -12387 (35) & 4 (4, 0) & s16\_0.4\_L2H & -11294 & -10712 (4) & 5 (5, 0)\\
			s14L\_0.4\_H2 & -11218 & -12294 (128) & 8 (7, 1) & s16\_0.4\_H2L & -9845 & -10676 (40) & 8 (8, 0)\\
			s14\_0.4\_H2H & -13621 & -12418 (4) & 5 (5, 0) & s16\_0.4\_H2H & -11579 & -10674 (42) & 8 (7, 1)\\
			\midrule
			s14\_0.6\_L2L & -11620 & -12297 (125) & 7 (7, 0) & s16\_0.6\_L2L & -8463 & -10563 (153) & 11 (10, 1)\\
			s14\_0.6\_L2H & -13620 & -12416 (6) & 8 (7, 1) & s16\_0.6\_L2H & -12161 & -10709 (7) & 10 (9, 1)\\
			s14\_0.6\_H2L & -10418 & -12297 (125) & 5 (4, 1) & s16\_0.6\_H2L & -10136 & -10677 (39) & 8 (7, 1)\\
			s14\_0.6\_H2H & -14020 & -12294 (128) & 6 (6, 0) & s16\_0.6\_H2H & -11581 & -10676 (40) & 4 (3, 1)\\
			\midrule
			s14\_0.8\_L2L & -11619 & -12139 (287) & 8 (7, 1) & s16\_0.8\_L2L & -8465 & -10563 (153) & 6 (5, 1)\\
			s14\_0.8\_L2H & -14418 & -12413 (9) & 9 (8, 1) & s16\_0.8\_L2H & -12108 & -10644 (52) & 14 (11, 3)\\
			s14\_0.8\_H2L & -10819 & -11925 (497) & 8 (7, 1) & s16\_0.8\_H2L & -10138 & -10569 (147) & 5 (5, 0)\\
			s14\_0.8\_H2H & -14413 & -12287 (135) & 7 (5, 2) & s16\_0.8\_H2H & -11582 & -10710 (6) & 7 (7, 0)\\               
			\bottomrule[1.25pt]
		\end{tabular}
	}
	\label{tab:benchmark}
	\vspace{-4mm}
\end{table}

Systematically, knowledge transfer is materialized in our experimentation by taking a \textit{descendant}-instance as the source-task and a \textit{parent}-instance as the target-task to be solved. That is, the goal is to solve a \textit{parent}-instance leveraging the knowledge of a previously solved \textit{descendant}-instance. 

First, all instances have been independently executed 10 times using D-Wave's \texttt{Advantage\_system6.4} by means of the common forward annealing process. Regarding the parameters, the number of runs was set to 1000. Furthermore, the \texttt{Knapsack} class belonging to the Qiskit v0.6.0 \textit{Optimization Applications} open library was employed for generating the KP QUBOs\footnote{\url{https://qiskit-community.github.io/qiskit-optimization/stubs/qiskit_optimization.applications.Knapsack.html}}. Thus, Table \ref{tab:benchmark} includes:
\begin{itemize}
	\item The best energy found for each instance across the 10 independent runs.
	\item The energy of the \textit{parent}-instance objective function if the best solution of the \textit{descendant}-instance would be applied, and the nominal value of the difference with respect to the energy of the best solution of the \textit{parent}-instance. For example, let's take the row related to \texttt{s16\_0.2\_L2L}; the best solution found in the \texttt{s16\_0.2\_L2L} resolution corresponds to an energy of -10133. This solution, when applied to the objective function of the \textit{parent}-instance, i.e., \texttt{s16}, produces an energy outcome of -10712. In turn, the energy of the best solution found for \texttt{s16} instance was -10716, which causes a difference of 4. 
	\item The Hamming distance between the best solutions of \textit{parent}- and \textit{descendant}-instances when executed independently. In addition, the decomposition of the Hamming distance is added. We specify, among the total number of variables that differ ($h$), which ones represent an item ($n_h$) and which ones are slack variables ($s_h$). Taking the same example, the Hamming distance between the best solutions found for \texttt{s16\_0.2\_L2L} and \texttt{s16} is equal to 9, meaning that the bitstrings of these solutions differ in 9 bits, 7 of which correspond to variables representing items and 2 to slack variables. 
\end{itemize} 

With all this in mind, we now proceed to answer the two questions posed in the introduction.

\subsection{\textit{\textbf{RQ1}: Is RA a paradigm that can benefit from knowledge transfer between problem instances with similar characteristics?}}\label{sec:RQ1}

A second set of tests has been conducted for properly answering \textbf{RQ1}, consisting of solving each target-instance \texttt{s14} and \texttt{s16} by means of RA procedure and using as input solution the best solutions found for each \textit{descendant}-instance through the first tests summarized in Table \ref{tab:benchmark}.

A fixed \texttt{RA} \texttt{schedule} has been used for all the tests, [(0.0, 1.0), (2.5, 0.5), (102.5, 0.5), (102.75, 1.0)], which has been obtained through an empirical procedure performed in lab. More precisely, an experimentation was carried out in which 25 different \texttt{RA} \texttt{schedule} were tested. Minor adjustments were made to the $s_p$, $t_r$, $t_p$, and $t_q$ parameters. To assess the quality of the designed \texttt{RA} \texttt{schedule}, each one was used to solve the two target-instance 10 times each, using one random solution for the problem as input. The aforementioned schedule was selected due to its superior performance compared to the others.

For the sake of fair comparison, the \texttt{annealing\_time} of all the forward annealing processes has been adjusted to the \texttt{RA} \texttt{schedule} duration so that all runs carried out in this research access the quantum computer the same amount of time. We recommend papers such as \cite{grant2021benchmarking} to readers interested in the analysis of the reverse annealing schedule. Lastly, \texttt{reinitialize\_state=True}.

For each (source instance - target instance) combination, 10 independent runs have been executed, with Table \ref{tab:res} depicting the best result found among these executions, along with the average and the standard deviation. As an example, 10 independent runs have been executed to solve \texttt{s14}, using the best solution found for \texttt{s14\_0.2\_L2L} as the input bitstring in the RA process (i.e., source of knowledge). For this combination, the energy of the best solution found is -12421, while the average and standard deviation are -12420.2 and 0.74, respectively. For comparison purposes, we also depict in Table \ref{tab:res} the baseline results obtained by \texttt{s14} and \texttt{s16} by means of forward annealing. In these cases, the source of knowledge has been represented as ``--".  It is important to highlight that the optimal results for each instance have been achieved using the industry-oriented Quantagonia Hybrid Solver (QHS, \cite{osaba2024hybrid}). In essence, this algorithm is divided into two distinguishable steps: Initially, it executes a set of \textit{primal heuristics} to solve the entire problem. These heuristics can be classical or quantum, utilizing external services. Subsequently, \texttt{QHS} enhances the best solution identified by the primal heuristics through a classical Branch-and-Bound algorithm. One of the key strengths of \texttt{QHS} is the inclusion of an \textit{optimality proof} mechanism, which provides the \textit{optimality gap} along with the results. The \textit{optimality gap} indicates the remaining potential in the optimization process. This feature is the primary reason we selected \texttt{QHS} as the solver to achieve optimal results for the considered instances.

\begin{table}[!t]
	\centering
	\caption{Results of the Reverse Annealing experimentation. Optimal results,  obtained using the industry-oriented Quantagonia's Hybrid Solver: \texttt{s14}=12422; \texttt{s16}=10718. In \textbf{bold} results that equalize or improve the performance of the forward annealing baseline (which is -12422 for \texttt{s14} and -10716 for \texttt{s16}).}
	\resizebox{0.70\columnwidth}{!}{
		\begin{tabular}{c|c|c|c}
			\toprule[1.5pt]                
			\multicolumn{2}{c|}{\textbf{Performance analysis on s14}} & \multicolumn{2}{c}{\textbf{Performance analysis on s16}}\\
			\makecell[c]{Source of\\Knowledge} & (best, avg., st.) & \makecell[c]{Source of\\Knowledge} & (best, avg., st.) \\
			\toprule[1.0pt]
			-- & (-12422, -12418.9, 2.02) & -- & (-10716, -10712.0, 3.00)\\ \midrule
			s14\_0.2\_L2L & (-12421, \textbf{-12420.2}, \textbf{0.74}) & s16\_0.2\_L2L & (\textbf{-10716}, \textbf{-10714.7}, \textbf{1.27}) \\
			s14\_0.2\_L2H & (\textbf{-12422}, \textbf{-12421.6}, \textbf{0.48}) & s16\_0.2\_L2H & (\textbf{-10718}, \textbf{-10716.2}, \textbf{0.97}) \\
			s14\_0.2\_H2L & (\textbf{-12422}, \textbf{-12421.0}, \textbf{0.63}) & s16\_0.2\_H2L & (\textbf{-10718}, \textbf{-10716.6}, \textbf{0.80}) \\
			s14\_0.2\_H2H & (\textbf{-12422}, \textbf{-12420.8}, \textbf{0.60}) & s16\_0.2\_H2H & (-10714, \textbf{-10711.5}, \textbf{1.43}) \\
			\midrule
			s14\_0.4\_L2L & (-12421, \textbf{-12420.6,} \textbf{0.48}) & s16\_0.4\_L2L & (\textbf{-10717}, \textbf{-10715.7}, \textbf{0.90}) \\
			s14\_0.4\_L2H & (\textbf{-12422}, \textbf{-12420.8}, \textbf{0.60}) & s16\_0.4\_L2H & (\textbf{-10717}, \textbf{-10716.0}, \textbf{1.00}) \\
			s14\_0.4\_H2L & (-12421, \textbf{-12419.4}, \textbf{1.11}) & s16\_0.4\_H2L & (\textbf{-10716}, \textbf{-10714.8}, \textbf{0.87}) \\
			s14\_0.4\_H2H & (\textbf{-12422}, -12420.6, \textbf{0.80}) & s16\_0.4\_H2H & (\textbf{-10717}, \textbf{-10714.9}, \textbf{1.22}) \\
			\midrule
			s14\_0.6\_L2L & (-12421, \textbf{-12420.7}, \textbf{0.45}) & s16\_0.6\_L2L & (-10715, \textbf{-10713.8}, \textbf{0.74}) \\
			s14\_0.6\_L2H & (-12421, \textbf{-12418.9}, \textbf{1.04}) & s16\_0.6\_L2H & (-10715, \textbf{-10713.9}, \textbf{0.94})\\
			s14\_0.6\_H2L & (\textbf{-12422}, -12420.6, \textbf{0.66}) & s16\_0.6\_H2L & (\textbf{-10716}, \textbf{-10715.1}, \textbf{0.94}) \\
			s14\_0.6\_H2H & (-12421, \textbf{-12420.4}, \textbf{0.66}) & s16\_0.6\_H2H & (\textbf{-10718}, \textbf{-10715.8}, \textbf{0.87}) \\
			\midrule
			s14\_0.8\_L2L & (-12421, \textbf{-12419.7}, \textbf{1.10}) & s16\_0.8\_L2L & (\textbf{-10718}, \textbf{-10715.0}, \textbf{1.34}) \\
			s14\_0.8\_L2H & (-12420, \textbf{-12419.1}, \textbf{1.30}) & s16\_0.8\_L2H & (-10714, -10710.7, 1.55) \\
			s14\_0.8\_H2L & (-12421, \textbf{-12419.5}, \textbf{1.02}) & s16\_0.8\_H2L & (\textbf{-10717}, \textbf{-10716.5}, \textbf{0.50}) \\
			s14\_0.8\_H2H & (-12421, \textbf{-12420.2}, \textbf{0.60)} & s16\_0.8\_H2H & (\textbf{-10717}, \textbf{-10715.6}, \textbf{0.79}) \\             
			\bottomrule[1.25pt]
		\end{tabular}
	}
	\label{tab:res}
	\vspace{-5mm}
\end{table}

The results shown in Table \ref{tab:res} provide answers to \textbf{RQ1}. If we pay attention to \texttt{s14}, we could preliminarily conclude that the sharing of knowledge does not imply an improvement with respect to the best solution found by the annealer. This reflection cannot be replicated if we look at the values relative to the mean and standard deviation, where the use of solutions from \textit{descendant}-instances clearly implies better performance. In summary, the sharing of knowledge through RA leads to more reliable and robust behavior of the annealer in the case of \texttt{s14}.

The results are even more interesting in the case of \texttt{s16}. In this case, the results offered by the forward annealing process are considerably worse as a consequence of facing a larger and more complex problem. In this specific case, the ToK between problems stands as a mechanism that directly implies superior performance for the annealer. More specifically, the use of RA under the criteria established in this research leads, in the vast majority of cases, to a more reliable and robust performance and even to an increase in the probability of obtaining the optimal solution to the \textit{target} problem.

In summary, in view of the results shown in Table \ref{tab:res}, we can conclude, at least preliminarily, that knowledge sharing through RA is a promising strategy for solving optimization problems. However, all that glitters is not gold, as the good performance shown in these tests reveals a clear line of investigation. This is because, despite better results, there is no clear pattern to shed light on the requirements that two problems have to meet to ensure productive knowledge sharing.

These varying results expose that \textbf{RQ1} requires inspecting further the reasons why some \textit{descendant}-instances produce better results than others. This inspection requires a more detailed examination of the relationship between the characteristics of the \textit{descendant}-instances and the performance shown in Table \ref{tab:res}.

\subsection{\textit{\textbf{RQ2}: Can we infer the characteristics that an input solution should meet to help increase the probability of succeeding in a RA process?}}

Although there is a widespread belief in the literature that RA is more effective when the initial state is ``\textit{close}'' to the ground state~\cite{passarelli2020reverse}, there is no exact definition for the term ``\textit{close}'' in this context. Some papers base this concept on the energetic proximity between the input solution and the ground state~\cite{ding2024effective}, while others suggest that using metrics such as the Hamming distance could be beneficial, contrasting with just choosing states that are low in energy, which may be far from the correct ground state in terms of the quantum annealing procedure \cite{pelofske2023mapping}.

\begin{figure}[t]
	\centering
	\includegraphics[width=0.57\linewidth]{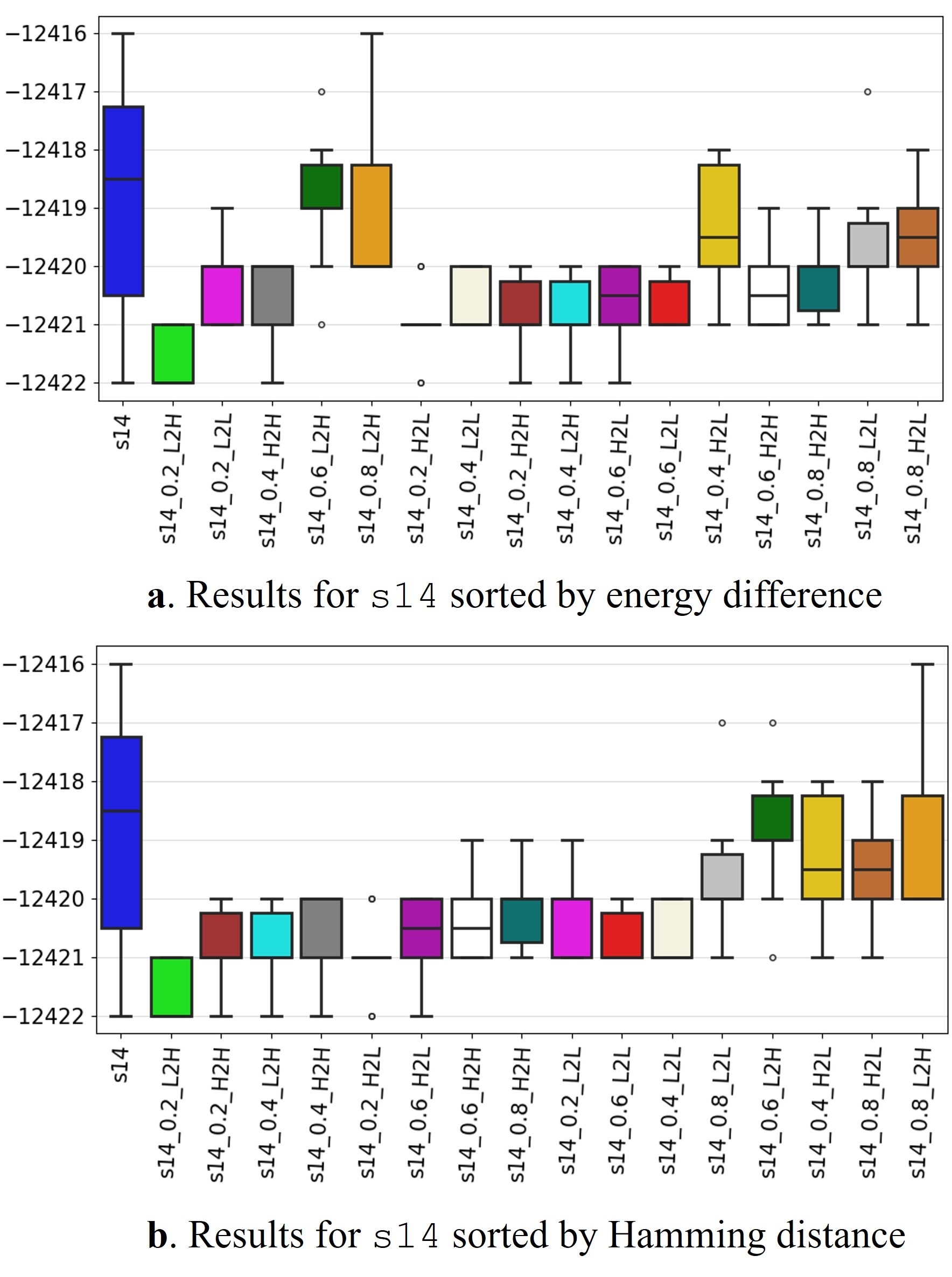}
	\caption{Results related to the resolution of \texttt{s14} using as input the best solution found for each \textit{descendant}-instance over 10 independent runs. The blue colored boxplots represent the baseline results obtained through forward annealing. The less the energy value, the better the solution.}
	\label{fig:s14}
\end{figure}

\begin{figure}[t]
	\centering
	\includegraphics[width=0.57\linewidth]{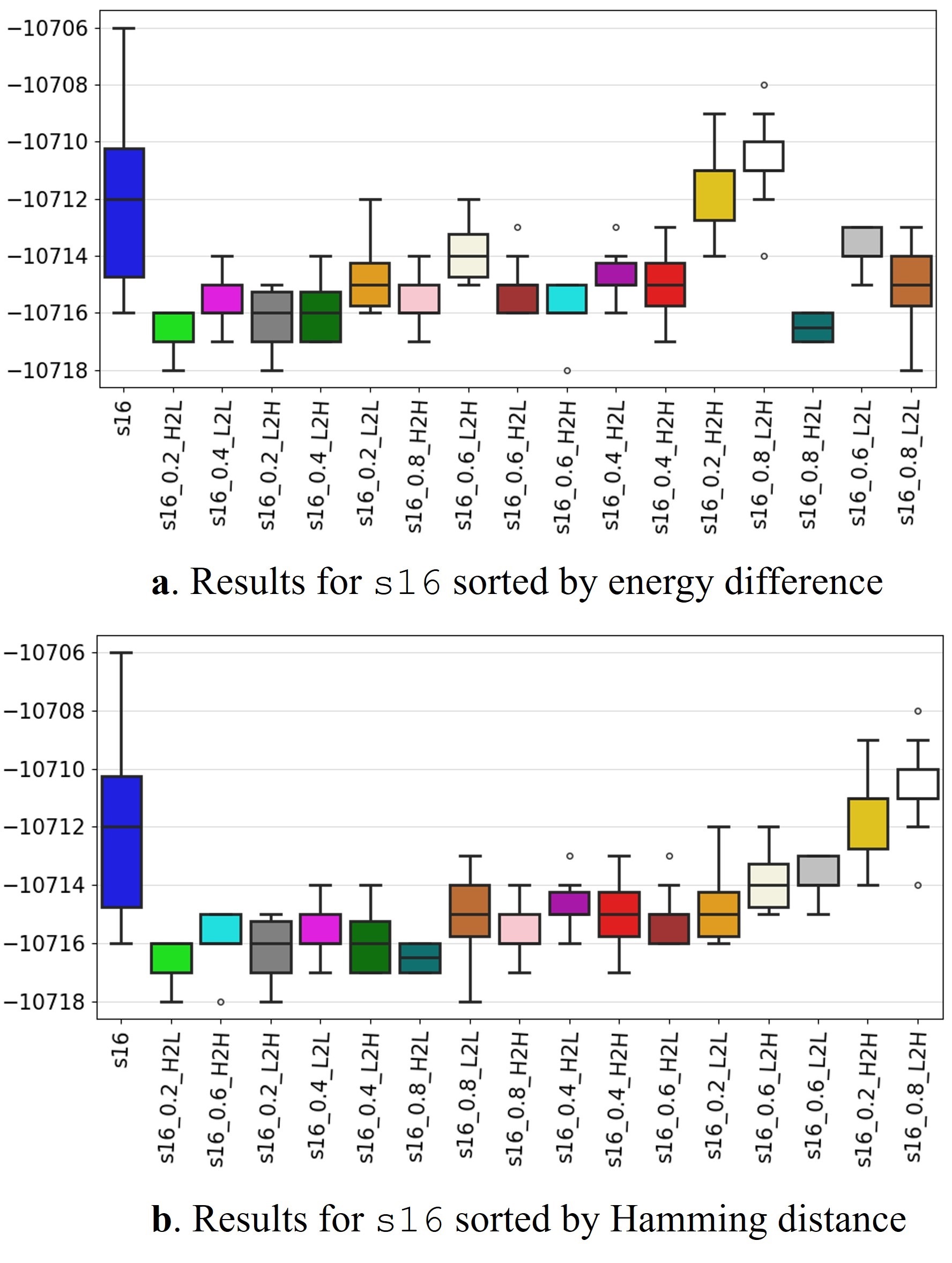}
	\caption{Results related to the resolution of \texttt{s16} using as input the best solution found for each \textit{descendant}-instance over 10 independent runs. The blue colored boxplots represent the baseline results obtained through forward annealing. The less the energy value, the better the solution.}
	\label{fig:s16}
\end{figure}

Thus, experimentally answering this second question can be done by sorting the results depicted in Table~\ref{tab:res}, using as references both energy differences and Hamming distances (represented in Table \ref{tab:benchmark}). This is the purpose of the plots shown in Figure \ref{fig:s14} and Figure \ref{fig:s16}. To facilitate the visualization of the outcomes, we show the results obtained by solving each target-instance through forward annealing in the leftmost part of these images. It should also be noted that the figures focused on the energy difference are based on the results shown in column \textit{Energy wrt. \texttt{sXX}} of Table \ref{tab:benchmark}. That is, on the energy value that each best-found solution returns using as a basis the objective function of its related target-instance. Also, the total Hamming distance $h$ has been used as a reference in Figure \ref{fig:s14}b and Figure \ref{fig:s16}b.

On the one hand, by carefully analyzing the results of Figure \ref{fig:s14}a and Figure \ref{fig:s16}a, it is difficult to detect a clear correlation between the energy difference among the solutions and their good performance as an input instance in an RA-based optimization process. Although the closest solutions offer competitive results, there are a large number of cases that make it impossible to draw a generalizable conclusion.

In the case of \texttt{s14}, \texttt{s14\_0.8\_L2H} has emerged as the worst instance for feeding the RA mechanisms, despite its energy value being close to the ground-state solution (-12413 vs. -12422, +9). On the contrary, \texttt{s14\_0.6\_H2L} has shown adequate performance despite its remoteness in terms of energy (-12297 vs. -12422, +125). Turning our attention to \texttt{s16}, it is also possible to detect that closer solutions produce good results. Even so, there are contradictory cases such as \texttt{s16\_0.6\_L2H}, \texttt{s16\_0.8\_H2L} or \texttt{s16\_0.8\_L2L}. Regarding \texttt{s16\_0.6\_L2H}, its performance has proven to be poor even being near the reference solution (-10709 vs. -10716, +7). Finally, the other two \textit{descendant}-instances mentioned have emerged as good alternatives to feed the RA, despite being among the most distant in terms of energy.

On the other hand, a clear trend is detected by analyzing Figure \ref{fig:s14}b and Figure \ref{fig:s16}b, where Hamming distance emerges as a key factor in finding good solutions for the target-instances. In other words, using as a source of knowledge a \textit{descendant}-instance whose best solution minimizes the Hamming distance regarding \textit{parent}-instances has proven to be efficient. Even a quick glance at the results obtained leads to the conclusion that as the Hamming distance increases, the performance of the RA degrades significantly.

This finding, despite being almost unexplored in studies related to RA, is in line with the premises of the classical computing field known as Transfer Optimization~\cite{gupta2017insights}. There are several studies related to this classical research stream that conclude that for efficient knowledge sharing, there must be, at least, a \textit{partial overlapping} in the optimal solutions to the problems at hand. More specifically, two problems present a \textit{partial overlapping} if the optima of both tasks are exactly the same in the unified search space regarding a subset of variables, and different in comparison to the rest of the variables \cite{da2017evolutionary}.

Furthermore, conclusions related to \textbf{RQ2} are also coherent with the work proposed in \cite{grass2019quantum}. That paper is focused on a specific kind of quantum annealing coined \textit{biased annealing}, which involves complementing the driver Hamiltonian with a longitudinal bias field. The importance of the Hamming distance between the driver Hamiltonian and the ground state is highlighted in a specific process coined \textit{iterative biased quantum annealing}, despite the same author minimizes its impact in a subsequent study when solving problems of large size \cite{grass2022quantum}.

Lastly, all the instances used in this paper and the complete set of results shown are openly available \cite{dataset}.

\section{Conclusions \& Further Work} \label{sec:conc}

In this work, we have preliminarily studied the influence of transferring knowledge between similar tasks through the reverse annealing mechanism implemented by D-Wave. To do that, two research questions have been posed:

\begin{quote}
	\textit{\textbf{RQ1}: Is RA a paradigm that can benefit from knowledge transfer between problem instances with similar characteristics?}
	
	\item \textit{\textbf{RQ2}: Can we infer the characteristics that an input solution should meet to help increase the probability of succeeding in a RA process?}     
\end{quote}

Firstly, answering \textbf{RQ1} and in view of the positive results obtained, the transfer of knowledge in quantum computing appears to be a promising research avenue that undoubtedly deserves further investigation. As for RA, it is commonly applied as a local refinement procedure right after an optimization process carried out by forward annealing or a classical optimization technique. In this work, RA serves as a mechanism for knowledge transfer, allowing the reuse of a solution obtained in an independent optimization process.

It is important to highlight that, aside from the results regarding performance, knowledge sharing pursues savings in the use of computational resources. These savings are mainly materialized by resorting to previous outcomes 
when solving a new task under the assumption that industries are likely to face tasks that have a lot in common with each other
. This is especially useful in real-world environments where companies accumulate results from past planning exercises and continually receive new tasks that are sometimes not very distinct from those already executed.

Secondly, answering \textbf{RQ2} and according to the results depicted, the closeness in terms of energy is not related to the performance of RA, and only the closeness with respect to the Hamming distance is. This means that the neighborhood must be based on solution codification and not energy, making it a priority to organize the coding of the target problem so that it fits as much as possible with the source problem through a unified search space \cite{gupta2015multifactorial}.

In light of these positive results, there are several open research questions that should be further studied in subsequent investigations, which can be summarized as: 
\begin{itemize}
	\item[$\blacksquare$] Future work of considerable importance will likely focus on advancing towards the generalization of the conclusions drawn from this study. To reach this ambitious goal, the following steps are essential:
	\begin{itemize}
		\item Inspired by the findings described in \cite{grass2022quantum}, further experiments should be carried out with larger instances, with the intention of determining whether the conclusions are replicable in such cases.
		\item Further study the impact of the \texttt{RA} \texttt{Schedules} on knowledge transfer. To do this, the preliminary study already described in Section \ref{sec:exp} will be revisited, and the knowledge transfer paradigm will be examined under the mentioned 25 \texttt{RA} \texttt{Schedules}.
		\item Carry out a thorough analysis of computational resources and time.
		\item Examine whether the findings from the experimentation can be replicated in other optimization problems, such as the Traveling Salesman Problem or the Bin Packing Problem.
		\item Study the potential limitations when applying RA to more real-world scenarios.
	\end{itemize}
	
	\item[$\blacksquare$] Fairly compare RA as a local refinement method (using the same instance for source- and target- tasks) with RA as a transfer knowledge procedure.
	
	\item[$\blacksquare$] Define a similarity measure for transfer of knowledge purposes. This means, given a potential source-task and its results, designing a metric that calculates how good a candidate source-task is for knowledge transfer purposes. The findings provided by the experimentation of this paper, particularly that focused on Hamming distance, are a good starting point for formulating this metric. Thus, a validated and accepted-by-the-community metric would be a milestone for the transfer of knowledge research line.
	
\end{itemize}

\section*{Conflict of Interest Statement}
The authors declare that the research was conducted in the absence of any commercial or financial relationships that could be construed as a potential conflict of interest.

\section*{Author Contributions}

E.O. and E.V.R. conceived the research and the experiments. E.O. and E.V.R developed the code and conducted the experimentation. E.O. and E.V.R. wrote and reviewed the manuscript.

\section*{Funding}
This work was supported by the Basque Government through Plan complementario comunicación cuántica (EXP. 2022/01341) (A/20220551). and by the Spanish CDTI through Misiones Ciencia e Innovación 2021 program (\textit{CUCO}, MIG-20211005).

\section*{Acknowledgments}
Authors thank Jordi Riu, Andrés Navas, and Josep Bosch from Qilimanjaro - Quantum Tech for the helpful discussions.

\section*{Data Availability Statement}
The datasets generated for this study and the complete set of results shown can be found in the \textit{Benchmark dataset and results for "Transfer of Knowledge through Reverse Annealing" experimentation} repository: \url{http://dx.doi.org/10.17632/yr8dg923wg.1}

\bibliographystyle{Frontiers-Vancouver}
\bibliography{biblio}

\end{document}